\documentclass[11pt]{article}

\begin{document}
\textheight 21 cm
\textwidth 13.4 cm

{\bf \Large Constraints on cosmological parameters through clusters XLF and XTF.}\\

{A. Del Popolo} \& N. Ercan\\
{Bo$\breve{g}azi$\c{c}i University, Physics Department,
     80815 Bebek, Istanbul, Turkey
}

\vskip 0.2cm
{\bf Abstract}\\
In this paper, I revisit the constraints obtained
by several authors (Reichart et al. 1999; Eke et al. 1998; Henry 2000)
on the estimated values of $\Omega_{\rm m}$, $n$ and $\sigma_8$ in the light of recent theoretical developments: 1) new theoretical mass functions (Sheth \& Tormen 1999, Sheth, Mo \& Tormen 2001, Del Popolo 2002b); 2) a more accurate mass-temperature relation, also determined for arbitrary $\Omega_{\rm m}$ and $\Omega_{\rm \Lambda}$ (Del Popolo 2002a). 

Firstly, using the quoted improvements, I re-derive an expression for the X-ray Luminosity Function (XLF), similarly to Reichart et al. (1999), and then I get some constraints to $\Omega_{\rm m}$ and $n$, by using the {\it ROSAT} BCS and {\rm EMSS} samples and maximum-likelihood analysis. Then I re-derive the X-ray Temperature Function (XTF), similarly to Henry (2000), re-obtaining the constraints on $\Omega_{\rm m}$, $n$, $\sigma_8$.
Both in the case of the XLF and XTF, the changes in the mass function and M-T relation produces an increase in $\Omega_{\rm m}$ of $ \simeq 20\%$ and similar results in $\sigma_8$ and $n$.
\vskip 0.1cm
{\bf Introduction}
\vskip 0.1cm
It is well known that clusters are strong X-ray emitters whose study
can put constraints on fundamental cosmological parameters.
There are different methods to trace the evolution of the cluster number density: 
a) The X-ray temperature function (XTF) has been presented for local (e.g. Henry \& Arnaud 1991) and distant clusters (Eke et al. 1998; Henry 2000). 
b) The evolution of the X-ray luminosity function (XLF).\\
The results obtained for $\Omega_{\rm m}$ and other cosmological parameters are in many cases discrepant the one with the other. Several studies in literature, show
that the parameters values span the entire range of acceptable solutions: $0.2 \leq \Omega_{\rm m} \leq 1$ (see Reichart et al. 1999).
The reasons leading to the quoted discrepancies has been studied in several papers (Eke et al. 1998; Borgani et al. 2001) 
( 1) The inadequate approximation given by the PS (e.g., Bryan \& Norman 1997). 2) Inadequacy in the structure formation as described by the spherical model leading to changes in the threshold parameter $\delta_{\rm c}$ (e.g., Governato et al. 1998). 3) Inadequacy in the M-T relation obtained from the virial theorem (see Del Popolo 2002a). 4) Effects of cowling flows. 5) Determination of the X-ray cluster catalog's selection function. 6) Evolution of the L-T relation. 7) Optimization methods used in the analysis.)
These reasons lead me to re-calculate the constraints on $\Omega_{\rm m}$, $n$ and $\sigma_8$, using the XLF and XTF.
\vskip 0.1cm
{\bf Constraints to $\Omega_{\rm m}$ and $n$ from the XLF}
\vskip 0.1cm
Similarly to Reichart et al. (1999), I re-derived an expression for the XLF, using an improved version of the mass function and M-T relation, obtained in Del Popolo 2000a, Del Popolo 2000b, respectively, taking account of the effects of asphericity and tidal interaction with neighbors. Then I got some constraints to $\Omega_{\rm m}$ and $n$, by using the {\it ROSAT} BCS and {\rm EMSS} samples.

As described in Del Popolo (2000b), the
mass function can be approximated by:
%
%
\begin{equation}
n(m,z)\simeq 1.21 \frac{\overline{\rho}}{m^{2}}\frac{d\log (\nu )}{d\log m}
\left( 1+\frac{0.06}{\left( a\nu \right) ^{0.585}}\right) \sqrt{\frac{a\nu }{2\pi }}\exp{\{-a\nu \left[ 1+\frac{0.57}{\left( a\nu \right) ^{0.585}}\right] ^{2}/2\}}
\label{eq:nmm}
\end{equation}
where $a=0.707$.
Eq. (\ref{eq:nmm}) can be converted from a mass function to a luminosity function using a L-T relation (I use that of  
Mathiesen \& Evrard (1998)), 
and a T-M relation. This last is the one obtained in Del Popolo (2000a), and is based on the merging-halo formalism of Lacey \& Cole (1993), accounting for the fact that
massive clusters accrete matter quasi-continuously, and again take account of angular momentum acquisition by protostructures:
\begin{equation}
kT \simeq 8 keV \left(\frac{M^{\frac 23}}{10^{15}h^{-1} M_{\odot}}\right)
\frac{
\left[
\frac{1}{m_1}+\left( \frac{t_\Omega }t\right) ^{\frac 23}
+\frac{K(m_1,x)}{M^{8/3}}
\right]
}
{
\left[
\frac{1}{m_1}+\left( \frac{t_\Omega }{t_{0}}\right) ^{\frac 23}
 +\frac{K_0(m_1,x)}{M_{0}^{8/3}}
\right]
}
\label{eq:kTT1}
\end{equation}
(see Del Popolo 2000a for a derivation of the previous equation and the definition of the terms in it).
The luminosity function is obtained as in Reichart et al. (1999), using the mass function and the M-T relation previously introduced (see Del Popolo 2003 for a detailed analysis).
A Bayesian inference analysis used to constraint the model parameters  
shows that: $\Omega_{\rm m}=1.15^{+0.40}_{-0.33}$ and
$n=-1.55^{+0.42}_{-0.41}$.
The previous result shows that the change in the mass function and M-T relation gives rise to an increase of $\Omega_{\rm m}$ and $n$ of $\simeq 20\%$ with respect to Reichart's results.  
The lesson from the previous calculation is that taking account of non-sphericity in collapse and the fact that
massive clusters accrete matter quasi-continuously gives rise to a noteworthy change in the prediction of cosmological parameters, as $\Omega_{\rm m}$. In order to check the previous trend,
I have also estimated the value of $\Omega_{\rm m }$ following Borgani et al. (2001).
Analyzing the ROSAT Deep Cluster Survey (RDCS) and using the XLF to obtain constraints on cosmological parameters, Borgani et al. (2001), found that $\Omega_{\rm m}=0.35^{+0.13}_{-0.10}$. 
Using their method and data, but our mass function and M-T relation, one obtains larger values of $\Omega_{\rm m}$ ($\Omega_{\rm m} \simeq 0.4 \pm 0.1$) that differently from the previous analysis (Reichart et al. 1999) exclude an Einstein-de Sitter model.
\vskip 0.1cm
{\bf Constraints to $\Omega_{\rm m}$, $n$, and $\sigma_8$ from the XTF}
\vskip 0.1cm
As previously reported, the mass function (MF) is a critical ingredient in putting strong constraints on
cosmological parameters (e.g., $\Omega_{\rm m}$).
In the following, I'll re-calculate the constraints obtained by Henry (2000), by using the mass function and the M-T relation modified as described in the previous section Eq. (\ref{eq:nmm}) and Eq. (\ref{eq:kTT1}) (see Del Popolo 2003).
I use a maximum likelihood fit to the unbinned data in order to determine various model parameters.
The method is described in Marshall et al (1983). The likelihood function is given by their Eq. (2),
adapted to our present situation.
At this point, we can fit the data described in Section. 2 of Henry (2000) to the theory previously described using
the quoted maximum likelihood method.
The most general description of the results requires the three parameters of the fit ($\Omega_{\rm m}$, $\sigma_8$ and n). 
These values shows that the correction introduced by the new form of the mass function and M-T relation gives rise to higher values of $\Omega_{\rm m}$ ($\Omega_{\rm m}=0.6 \pm 0.13$, while it is $\Omega_{\rm m}=0.49 \pm 0.12$ for Henry (2000)) and
$n=-1.5 \pm 0.32$ ($n=-1.72 \pm 0.34$ in Henry (2000)).
Constraints are relatively tight when
considering this single parameter. We find that $\Omega_{\rm m}=0.6^{+0.12}_{-0.11}$
at the 68\% confidence level and $\Omega_{\rm m}=0.6^{+0.23}_{-0.2}$ 
at the 95\% confidence level for the open model.

Concluding, our analysis shows that improvements in the mass function and M-T relation increases the value of $\Omega_{\rm m}$ and that even small correction in the physics of the collapse can induce noteworthy effects on the constraints obtained. 

{\bf References}
\vskip 0.2cm
Borgani S., et al. 
2001, ApJ 561, 13\\
Bryan G.L., Norman M.L., 1997, ApJ 495, 80\\
Del Popolo A., Gambera M., 1998, A\&A 337, 96\\
Del Popolo A., 2002a, MNRAS 336, 81\\
Del Popolo A., 2002b, MNRAS 337, 529\\
Del Popolo A., 2003,ApJ 599, 723\\ 
Eke V. R., Cole S, Frenk C. S., Henry J.P., 1998, MNRAS 298, 1145\\
Governato F., et al., 
1999, MNRAS 307, 949\\
Henry J.P., 2000, ApJ 534, 565\\
Lacey C., Cole S., 1993, MNRAS 262, 627\\
Marshall H. L., et al., 
1983, ApJ 269, 35\\
Reichart D. E., et al., 
1999, ApJ 518, 521\\
Sheth R. K., Tormen G., 1999, MNRAS 308, 119\\
Sheth R. K., Mo H. J., Tormen G., 2001, MNRAS 323, 1 (SMT)\\

\end{document}